\documentclass[twocolumn,twocolappendix,tighten,times]{aastex7}

\usepackage{graphicx}
\usepackage{amsmath}
\usepackage{comment}

\shorttitle{An Unusually Clumpy UDG}
\shortauthors{Tang et al.}

\begin{document}

\title{An Unusually Clumpy Ultra-Diffuse Galaxy in the Coma Cluster: A Missing Link for Globular Cluster Formation?}

\correspondingauthor{Yimeng Tang}
\email{ymtang@ucsc.edu}

\author[0000-0003-2876-577X]{Yimeng Tang}
\affiliation{Department of Astronomy \& Astrophysics, University of California Santa Cruz, 1156 High Street, Santa Cruz, CA 95064, USA}
\email{ymtang@ucsc.edu}

\author[0000-0003-2473-0369]{Aaron J.\ Romanowsky}
\affiliation{Department of Astronomy \& Astrophysics, University of California Santa Cruz, 1156 High Street, Santa Cruz, CA 95064, USA}
\affiliation{Department of Physics \& Astronomy, San Jos\'e State University, One Washington Square, San Jose, CA 95192, USA}
\email{aaron.romanowsky@sjsu.edu}

\author[0000-0003-2898-0728]{Nobuhiro Okabe}
\affiliation{Physics Program, Graduate School of Advanced Science and Engineering, Hiroshima University, 1-3-1 Kagamiyama, Higashi-Hiroshima, Hiroshima 739-8526, Japan}
\email{okabe@hiroshima-u.ac.jp}

\author[0000-0001-9742-3138]{Kevin A.\ Bundy}
\affiliation{Department of Astronomy \& Astrophysics, University of California Santa Cruz, 1156 High Street, Santa Cruz, CA 95064, USA}
\email{kbundy@ucsc.edu}

\begin{abstract}

We present a study of Coma-B122, an unusually clumpy star-forming ultra-diffuse galaxy (UDG) in the Coma cluster outskirts, using deep Keck Cosmic Web Imager integral-field spectroscopy and Subaru Hyper Suprime-Cam imaging. B122 hosts $\sim10$ luminous star-forming clumps contributing $\sim25$\% of its luminosity and $\sim7$\% of its stellar mass, with typical clump masses of $\sim10^6\ M_\odot$ comparable to the most massive super star clusters known in nearby starbursting dwarfs. The clumps form a distinctive ring configuration $\sim2$ kpc from the center. Full-spectrum fitting shows that $\sim12.5$\% of the diffuse body's stellar mass formed during an extended episode of enhanced star formation over the past $\sim300$ Myr, within which all the clumps assembled. B122 shows evidence of weak ram-pressure stripping (RPS), including a diffuse light distribution truncated on the side facing the Coma center and extended to the opposite side, along with an asymmetric H~{\sc i} velocity profile. Its ordered stellar and ionized-gas rotation, cold gas fraction, and phase-space position are consistent with an early stage of infall. Spatial variations in clump star formation histories and gas-phase abundances further indicate that RPS-driven metal-poor gas inflow may have triggered and sustained the recent star formation. B122 may offer a rare present-day view of how UDGs assemble their globular cluster (GC) systems during cluster infall, although it remains uncertain if the clumps could survive as GCs. Future observations with higher spatial resolution and of more analogs are important to provide a complete picture of the formation and evolution of clumpy low-surface-brightness dwarf galaxies.

\end{abstract}


\section{Introduction} \label{sec:intro}

Ultra-diffuse galaxies (UDGs) are low-surface-brightness (LSB) systems with dwarf-like luminosities but sizes comparable to those of giant galaxies, typically defined by central surface brightness $\mu_{g,0}>24$ mag arcsec$^{-2}$ and effective radius $R_{\rm e}>1.5$ kpc \citep{vanDokkum2015}. A variety of mechanisms have been proposed to produce such ``puffy'' dwarf galaxies, including tidal heating, stellar feedback, and high gas angular momentum (e.g., \citealt{Rong2017,Jiang2019,Sales2020,Benavides2023}). However, current cosmological simulations incorporating these processes still struggle to reproduce the full observed diversity of dwarf galaxy sizes, morphologies, and dynamical structures, pointing to missing physics in our understanding of dwarf galaxy formation \citep{Sales2022,Martin2025}. 

A tantalizing clue comes from the widespread existence of UDGs with populous systems of globular clusters (GCs; \citealt{Beasley2016,vanDokkum2017,Lim2018,Forbes2020}). These systems pose an apparent paradox: massive bound clusters form most efficiently in high-pressure, intensely star-forming environments \citep{Kruijssen2012,Adamo2020}, whereas their present-day hosts have exceptionally diffuse stellar bodies.

The most conventional explanation for the GC-rich UDGs is that they were originally dense, turbulent dwarf galaxies that expanded through tidal heating after early infall into a galaxy cluster \citep{Carleton2021,Pfeffer2024}. Indeed, most isolated UDGs are observed to host few if any GCs \citep{Jones2023}, and the vast majority of known GC-rich UDGs are in groups and clusters. However, some GC-rich UDGs have been found in late-infall regions of clusters, and in other low-density environments -- implying another, unconventional formation mechanism (e.g., \citealt{vanDokkum2019,Gannon2022,Janssens2022,Fielder2023,Tang2026a,Gannon2026}). 

One possible pathway is the ``failed galaxy'' scenario, in which an early, intense, cluster-dominated burst is followed by rapid quenching, leaving the galaxy GC-rich but underluminous for its halo \citep{Danieli2022}. This scenario may connect to the clump-dominated mode of star formation observed in high-redshift galaxies with the James Webb Space Telescope (e.g., \citealt{Chen2023,Adamo2024}). Another possibility is a high-speed collision of gas-rich dwarfs that naturally produced dense star clusters within a diffuse galaxy, leading also to an unusually top-heavy GC mass function and a lack of dark matter \citep{Silk2019,Lee2021,vanDokkum2022}. 

Direct observational tests of such scenarios are very difficult since they would have occurred at high redshifts, leading to old UDGs and GCs today \citep{FerreMateu2023}. However, such processes may continue at a reduced rate at later times, presenting the goal of finding any nearby UDGs with young massive clusters (YMCs) or dense star-forming clumps that could evolve into GCs. Clumpy star formation occurs in both compact and diffuse dwarfs \citep{Elmegreen2012,James2015,James2017}, and several isolated UDGs hosting YMCs have recently been identified \citep{Buzzo2025c,Buzzo2025d,Khim2025}. Such clumps may be sustained by gas accretion \citep{Lassen2025,Long2025}. However, star-forming UDGs generally have low star formation rate (SFR) local surface densities \citep{Kado-Fong2022a,Kado-Fong2022b}, suggesting that the formation of massive bound clusters may require additional mechanisms to compress or concentrate the gas. Such mechanisms may be particularly effective in the vicinity of the most massive clusters like Coma and Perseus (virial mass $M_{\rm vir} \sim 10^{15} M_\odot$), which preferentially host more GC-rich UDGs (see \citealt{Lim2020}).

\begin{figure}
\centering
\includegraphics[width=0.472\textwidth]{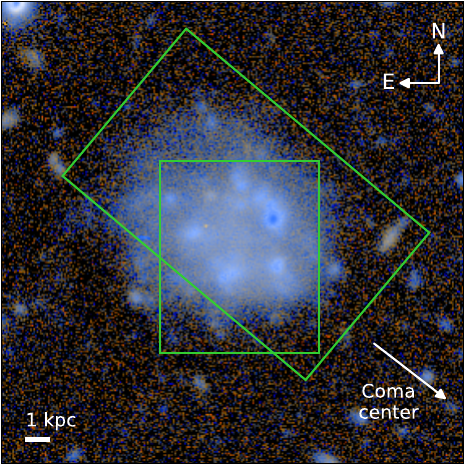}
\caption{Subaru/HSC pseudo-color image of Coma-B122 produced using the $g$ and $r$ bands. The 1 kpc scale bar indicates the physical size. Some bright clumps can be clearly seen within the galaxy. The green rectangles show the field of view of the Keck/KCWI observations. The white arrow indicates the projected direction toward the Coma cluster center.}
\label{fig:b122_hsc_color_image}
\end{figure}

Here we present a highly unusual star-forming UDG on the outskirts of the Coma cluster ($R_{\rm proj} \approx 0.8\ R_{\rm vir}$), Coma-B122 (AGC~234238\footnote{Other identifications for B122 in the literature include GMP~0521 \citep{Godwin1983}, NSA 142918 \citep{Blanton2011}, and SMDG J1304223+284552 \citep{Zaritsky2023b}. Its first spectroscopic redshift was obtained by \cite{Salzer2001}.}; shown in Figure~\ref{fig:b122_hsc_color_image}), providing an ideal extreme case for investigating this question. B122 was initially cataloged as a UDG in the Subaru Hyper Suprime-Cam (HSC; \citealt{Miyazaki2018,Aihara2018}) Coma cluster survey of \cite{Bautista2023}. This galaxy has a confirmed redshift as a cluster member from the Dark Energy Spectroscopic Instrument (DESI; \citealt{DESI2026}). While many star-forming UDGs can be found in the infall regions of groups and clusters, B122 is unique among all such known objects in hosting a collection of $\sim$ 10 luminous star-forming clumps that are potential progenitors for GCs. Their ring-like configuration $\sim$~2~kpc from the galaxy center is reminiscent of a ring of unusually massive GCs in a nearby LSB dwarf FCC~224 \citep{Tang2025b}, while no comparably luminous clumps are found either at the center of B122 or in its outer regions. B122 suggests that the processes forming GC-rich UDGs may continue up to the present day, and it thereby provides a unique window for further insights into this unsolved problem.

Two broad scenarios could explain B122's unusual combination of diffuse stellar body and clumpy star formation. Ram pressure stripping (RPS) during its infall into Coma could compress the gas, triggering enhanced star formation before quenching (e.g., \citealt{Grishin2026}). B122's diffuse stellar light shows a striking asymmetry, with an extension toward the northeast, opposite the projected direction toward the Coma cluster center, and a truncation toward the southwest (Figure~\ref{fig:b122_hsc_color_image}) -- potentially associated with RPS. Alternatively, a dwarf--dwarf interaction or collision could create a similar starburst independent of the cluster environment \citep{Silk2019,Buzzo2025c,Buzzo2025d}. The two scenarios predict distinct patterns of star formation history (SFH) and kinematics, allowing them to be tested observationally.

In this work, we make use of new integral-field spectroscopy of B122 from the Keck Cosmic Web Imager (KCWI; \citealt{Morrissey2018}) on the Keck II telescope to investigate B122, especially the triggering mechanism of its star formation and bright clumps. We organize this paper as follows. In Section \ref{sec:data_methods}, we describe the Subaru/HSC imaging and Keck/KCWI spectroscopy of B122, as well as our data analysis methods. We present our results in Section \ref{sec:results} and discuss various evolutionary scenarios in Section \ref{sec:discussion}. A summary follows in Section \ref{sec:conclusion}. The adopted distance of B122 is the Coma cluster distance of 100 Mpc. All magnitudes are on the AB system, and the adopted Galactic extinction is 0.04 and 0.03 mag in the HSC $g$ and $r$-bands, respectively.

\section{Data \& Methods} \label{sec:data_methods}

\subsection{Photometric data and modeling}
\label{sec:galfit}

\begin{figure*}
\centering
\includegraphics[width=1\textwidth]{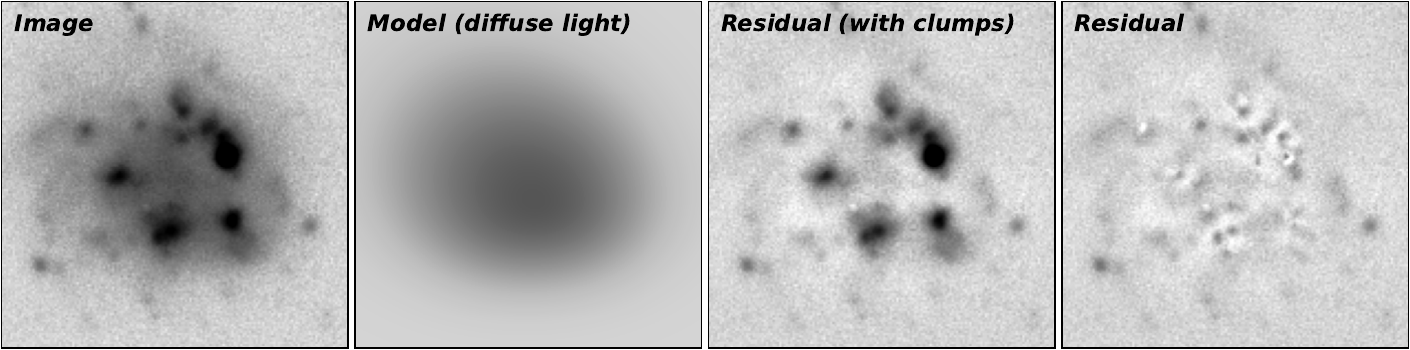}
\caption{B122 HSC $g$-band image as an example of our GALFIT modeling. From left to right, the four panels show the original image, the diffuse galaxy model, and the residuals (with and without clump components). The images shown here have sizes of 4 times the major-axis $R_{\rm e}$ of B122 ($4R_{\rm e}=26.5''$, or 12.8 kpc). North is up, and East is left.}
\label{fig:galfit}
\end{figure*}

\begin{figure*}
\centering
\includegraphics[width=1\textwidth]{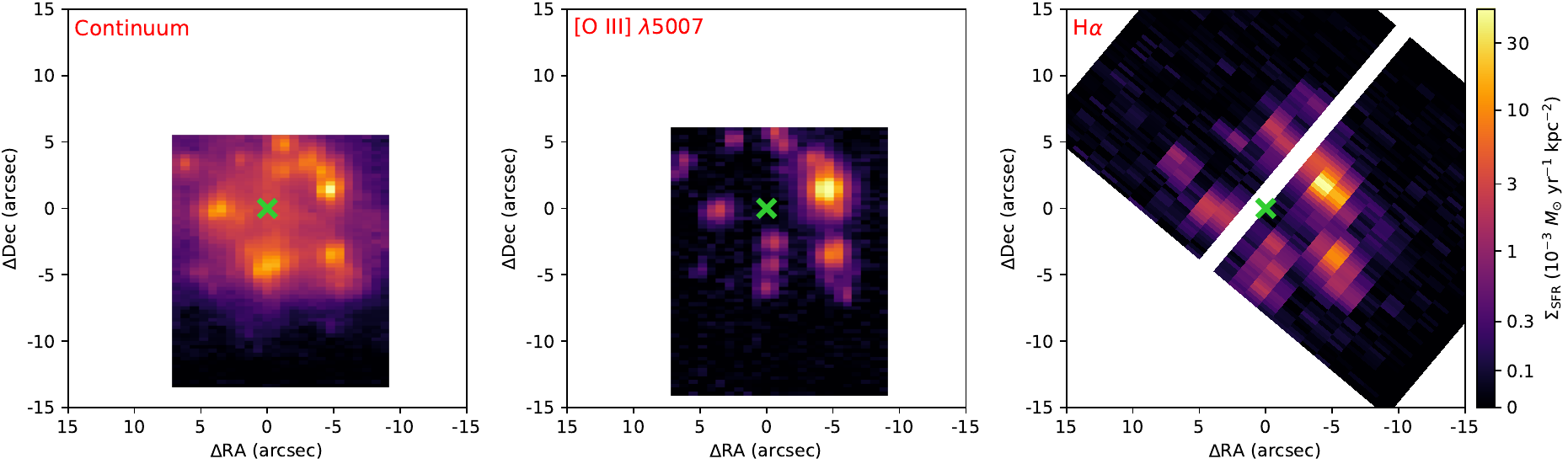}
\caption{The flux map of B122's continuum ({\em left}), and [O {\sc iii}] $\lambda$5007 ({\em middle}) and H$\alpha$ ({\em right}) emission lines, all shown on inverse hyperbolic sine (asinh) color scales. The colorbar of the H$\alpha$ panel gives the SFR surface density $\Sigma_{\rm SFR}$ derived from the H$\alpha$ flux. The spatial scale is 14.5$\times$14.5 kpc. There is wide variation in emission line fluxes among the stellar clumps with similar continuum level. Some stellar clumps seen in the continuum map have weak or undetected ionized-gas emission, while some ionized-gas knots in [O {\sc iii}] and H$\alpha$ map lack significant continuum counterparts.
}
\label{fig:luminosity_map}
\end{figure*}

\begin{figure*}
\centering
\includegraphics[width=1\textwidth]{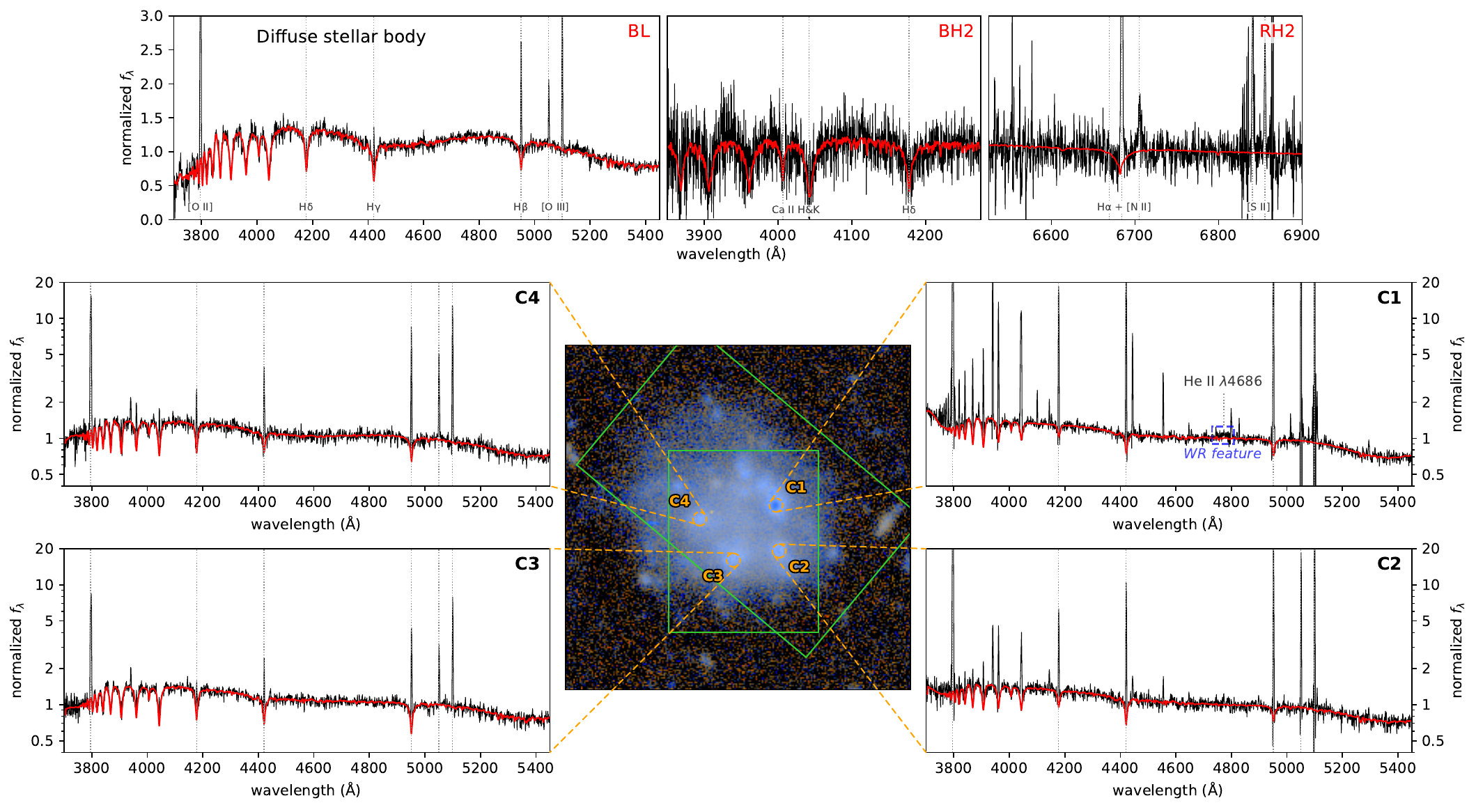}
\caption{Extracted spectra (black) of B122's diffuse stellar body ({\em top}), and the four brightest clumps ({\em left and right}). The best-fit stellar continuum is shown in red. For the diffuse stellar body, the three panels show the BL, BH2, and RH2 spectra, respectively. Only the BL spectra are shown for the clumps. The clump spectra show substantial variations in emission line strength relative to the stellar continuum, indicating spatial differences in gas conditions and recent star formation activity. Clump C1 also shows a Wolf--Rayet signature ``bump'' near He~{\sc ii} $\lambda4686$.}
\label{fig:spectra}
\end{figure*}

\begin{deluxetable}{lc}
\tabletypesize{\small}
\tablecolumns{2}
\tablehead{
\colhead{Properties} 
&\colhead{Value}
}
\startdata
RA (J2000) [deg] & 196.09277\\
Dec (J2000) [deg] & 28.76435\\
adopted distance [Mpc] & 100 \\
$R_{\rm e}$ [arcsec] & $6.61 \pm 0.11$ \\
$R_{\rm e}$ [kpc] & $3.20 \pm 0.05$ \\
S\'ersic $n$ & $0.52 \pm 0.03$ \\
$b/a$ & $0.77 \pm 0.02$ \\
position angle [deg] & $81 \pm 3$ \\
$m_{g,0}$ [mag] & $18.21 \pm 0.02$ \\
$M_{g,0}$ [mag] & $-16.79 \pm 0.02$ \\
$(g-r)_0$ [mag] & $0.16 \pm 0.02$ \\
$\mu_{g,0}$ [mag arcsec${}^{-2}$] & $23.93 \pm 0.04$ \\
$\langle \mu_{g} \rangle_{\rm e}$ [mag arcsec${}^{-2}$] & $24.32 \pm 0.03$ \\
radial velocity (stars) [km~s$^{-1}$] & $5495 \pm 4$ \\
radial velocity (ionized gas) [km~s$^{-1}$] & $5507 \pm 2$ \\
radial velocity (H {\sc i}) [km~s$^{-1}$] & $5510 \pm 5$ \\
log $M_\star/M_{\odot}$  & $8.02 \pm 0.10$ \\
log $M_{\rm HI}/M_{\odot}$  & $8.95 \pm 0.05$ \\
stellar age [Gyr] & $0.87 \pm 0.20$ \\
stellar metallicity [M/H] [dex] & $-1.26 \pm 0.18$ \\
SFR$_{\rm H\alpha}$ [$M_\odot\ {\rm yr^{-1}}$] & $0.052 \pm 0.005$ \\
SFR$_{\rm NUV}$ [$M_\odot\ {\rm yr^{-1}}$] & $0.063 \pm 0.006$ \\
12+log(O/H) & $8.1 \pm 0.1$ \\
$M_{\rm clump}/M_\star$  & $\sim 7\%$ \\
$L_{{\rm clump}, g}/L_g$  & $\sim 25\%$ \\
\enddata

\caption{Various properties of B122, including sky coordinates, adopted distance, $g$-band major-axis effective radius $R_{\rm e}$ in arcsec and kpc, S\'ersic index, axis ratio, photometric position angle, $g$-band total magnitude, $(g-r)_0$ color, central surface brightness in the $g$-band $\mu_{g,0}$ and average surface brightness $\langle \mu_{g} \rangle_{\rm e}$ of the diffuse stellar body, radial velocity of the diffuse stellar body, ionized gas, and H {\sc i} component, total stellar mass, H {\sc i} gas mass, mass-weighted stellar age and metallicity, total SFR estimated by H$\alpha$ flux and by NUV luminosity, gas-phase metallicity, and the stellar mass and $g$-band luminosity ratios between the clumps and the whole galaxy. Magnitudes and colors are corrected for Galactic extinction.}
\label{tab:properties}
\end{deluxetable}

The deepest imaging of B122 comes from Subaru/HSC observations of the Coma cluster, with the details presented in \cite{Bautista2023}. The exposure times for each of seven pointings are approximately 136 and 48 minutes in the $g$ and $r$ bands, respectively. The data were processed with \texttt{hscpipe} version 8 \citep{Bosch2018},
following the standard reduction procedure, including single-visit processing, mosaic construction, image coaddition, and photometric calibration against the Pan-STARRS1 Survey \citep{Schlafly2012}. The final coadded image has a seeing full-width at half-maximum of approximately $1\farcs0$.

\cite{Bautista2023} used HSC imaging with a source-cleaning procedure optimized for detecting diffuse galaxies in the presence of compact sources and extended stellar halos. Their survey covered the full Coma cluster to beyond $R_{\rm vir}$, approximately twice the area of previous Subaru studies, and identified 774 UDGs ($R_{{\rm e},r}\geq1.5$ kpc) and 729 sub-UDGs ($1.0\leq R_{{\rm e},r}<1.5$ kpc). This work provides a substantially larger and wider-area sample than earlier Coma UDG and LSB catalogs \citep{vanDokkum2015,Yagi2016,Alabi2020}.

To decompose the HSC imaging of B122 into its diffuse stellar body and embedded star-forming clumps, we use an iterative fitting procedure based on {\tt GALFIT} \citep{Peng2002}. We first fit the diffuse galaxy light with a single S\'ersic profile, adding low-order Fourier modes to account for the lopsided structure and arc-like features seen in Figure \ref{fig:b122_hsc_color_image}. We then identify significant, spatially compact positive residuals in the residual image using image segmentation and deblending with {\tt photutils} \citep{Bradley2025}, and interpret these residuals as previously unmodeled clumps. Each newly identified clump is modeled with a compact Gaussian profile and added to the {\tt GALFIT} model. The full model, including the galaxy and all previously identified clumps, is then refitted simultaneously, allowing all structural parameters to vary. This iteration is repeated until no significant residual structure remains or the reduced $\chi^2$ no longer improves. 

Our {\tt GALFIT} fitting for the HSC $g$-band imaging is shown in Figure \ref{fig:galfit}. The diffuse stellar body has a major-axis $R_{\rm e}=6.61\pm0.11$ arcsec (or $3.20\pm0.05$ kpc) and $\mu_{g,0}=23.93\pm0.04$ mag~arcsec$^{-2}$. Its total magnitude is $m_{g,0}=18.21\pm0.02$ mag, with a color of $(g-r)_0=0.16\pm0.02$ mag (including the clumps). B122 is also detected by GALEX, with a Galactic extinction corrected NUV magnitude of $m_{\rm NUV,0}=19.44 \pm 0.03$~mag ($A_{\rm NUV}=0.10$~mag) from the NASA/IPAC Extragalactic Database\footnote{\href{https://ned.ipac.caltech.edu}{https://ned.ipac.caltech.edu}}. Using the calibration of \citet{Kennicutt2012}, this flux translates to a global SFR of ${\rm SFR}_{\rm NUV}=0.063\pm0.006\ M_\odot\ {\rm yr^{-1}}$. These and other photometric measurements are summarized in Table~\ref{tab:properties}.

We detect individual clumps down to $M_{g,0} \sim -11$, with the brightest $\sim6$ clumps reaching luminosities of $M_{g,0} \sim -13$. We find that the clumps collectively contribute $\sim25\%$ of the total galaxy luminosity in $g$-band. The clumps appear broader than the HSC $g$-band point-spread function, suggesting that they are unlikely to be individual YMCs. However, blending with diffuse galaxy light and strong emission-line contamination prevent reliable size measurements. We do not find sources in B122 consistent with the properties of old, red GCs, as these would be fainter than our detection limit.

\subsection{Spectroscopic data and modeling}
\label{sec:ppxf}

Integral-field spectroscopy of B122 was obtained with Keck/KCWI on 2025 May 23 in program R372 (PI: Tang)\footnote{Shallower KCWI observations were taken with a similar set-up on 2025 Apr 30 (program U234, PI: J.~Brodie), but are not used here. Additional observations with the RH3 grating and Medium slicer were obtained on 2025 May 23 but are not used in the further analysis because of poor data quality.}. The seeing was $1\farcs0$. We observed B122 in several different configurations with various wavelength coverage and spectral resolution:

\begin{itemize}
\item BL grating with the Medium slicer: $1800\,{\rm s} + 4\times1200\,{\rm s}$ ($R \sim 1800$, wavelength coverage $\sim$ 3400--5600 \AA);
\item BH2 grating with the Large slicer: $3\times1200\,{\rm s}$ ($R \sim 4500$, wavelength coverage $\sim$ 3820--4320 \AA);
\item RH2 grating with the Large slicer: $580\,{\rm s} + 5\times540\,{\rm s}$ ($R \sim 4500$, wavelength coverage $\sim$ 6250--7150 \AA).
\end{itemize}

\begin{deluxetable*}{lcccccccc}
\tabletypesize{\footnotesize}
\tablecolumns{9}
\tablehead{
\colhead{Clump}
&\colhead{RA (J2000)}
&\colhead{Dec (J2000)}
&\colhead{$M_{g,0}$}
&\colhead{$(g-r)_0$}
&\colhead{radial velocity}
&\colhead{$\log M_\star/M_\odot$}
&\colhead{SFR$_{\rm H\alpha}$}
&\colhead{12+log(O/H)}\\
\colhead{}
&\colhead{[deg]}
&\colhead{[deg]}
&\colhead{[mag]}
&\colhead{[mag]}
&\colhead{[km~s$^{-1}$]}
&\colhead{}
&\colhead{[$10^{-3}\ M_\odot\ {\rm yr^{-1}}$]}
&\colhead{}
}

\startdata
C1 & 196.09157 & 28.76485 & $-13.81 \pm 0.01$ & $-0.47 \pm 0.01$ & $5505 \pm 1$ & $6.0 \pm 0.2$ & $25 \pm 3$ & $7.90 \pm 0.09$\\
C2 & 196.09145 & 28.76351 & $-13.30 \pm 0.01$ & $0.06 \pm 0.02$ & $5517 \pm 2$ & $6.1 \pm 0.2$ & $7 \pm 2$ & $8.10 \pm 0.09$\\
C3 & 196.09300 & 28.76327 & $-13.28 \pm 0.01$ & $-0.05 \pm 0.02$ & $5508 \pm 2$ & $6.1 \pm 0.1$ & $2 \pm 1$ & $8.14 \pm 0.09$\\
C4 & 196.09416 & 28.76444 & $-12.55 \pm 0.01$ & $0.01 \pm 0.02$ & $5488 \pm 2$ & $5.8 \pm 0.1$ & $3 \pm 1$ & $8.16 \pm 0.09$\\
\enddata

\caption{Properties of the four brightest clumps of B122 (labeled in Figure \ref{fig:spectra}), including sky coordinates, $g$-band absolute magnitude and $(g-r)_0$ color from the {\tt GALFIT} modeling, radial velocity of the ionized gas, stellar mass, SFR from the H$\alpha$ flux, and gas-phase metallicity. Magnitudes and colors are corrected for Galactic extinction.}
\label{tab:clumps}
\end{deluxetable*}

These data were reduced using the standard KCWI data reduction pipeline and subsequently cropped spatially and spectrally to exclude regions affected by poor data quality. We estimate and subtract the sky background using adjacent regions of the galaxy in the datacube as an on-chip sky spectrum. For the red arm, the Zurich Atmospheric Purge \citep{Soto2016} algorithm is additionally applied to the residual sky features in the on-chip sky region of the reduced datacube to characterize the sky emission more accurately (e.g., \citealt{Buzzo2025b}). 

Figure~\ref{fig:luminosity_map} shows the flux maps of B122's continuum, [O~{\sc iii}] $\lambda$5007, and H$\alpha$ emission --- with color-coding for the corresponding SFR surface density based on the calibration of \cite{Kennicutt2012}. The continuum flux is measured as the mean flux density over a 200 \AA\ window centered at 4700 \AA, where no major emission lines are present. For the [O {\sc iii}] and H$\alpha$ emission lines, we integrate the flux within a 20 \AA\ window centered on each line and subtract the continuum contribution estimated from nearby wavelength channels without emission lines. Most of the stellar clumps coincide with clumps in the ionized gas, although the correspondence is not exactly one-to-one: some stellar clumps have weak or undetected ionized-gas emission, while some ionized-gas knots lack significant continuum counterparts (which may be too faint for our detection limits). In the following analysis and discussion, we consider only the stellar clumps.

We then extract spectra from $1\farcs5$-diameter apertures centered on the four brightest clumps for further analysis. No additional local background subtraction is applied to these clump spectra, because the underlying diffuse galaxy light exhibits a substantial spatial variation, making it difficult to reliably estimate and subtract its contribution using a simple background annulus. We also extract the spectrum of the diffuse galaxy light by masking all the visible clumps, in the region where the two KCWI fields of view overlap. Figure \ref{fig:spectra} shows the extracted spectra of B122's diffuse light and clumps, with the clumps labeled C1--C4.

We use the Penalized Pixel-Fitting method ({\tt pPXF}; \citealt{Cappellari2004,Cappellari2017}) to model the stellar population and kinematics of B122 (see red curves in Figure~\ref{fig:spectra}). The stellar continuum is modeled as a linear combination of single stellar population (SSP) templates, while the emission lines are fitted with Gaussian templates. We use the BH2 and RH2 grating configurations to measure the kinematics of the stellar body and ionized gas, respectively. For the stellar kinematics, we adopt high-resolution ($R\sim10000$) SSP models based on the X-shooter Spectral Library \citep{Verro2022}, with Padova+00 isochrones \citep{Girardi2000} and the \cite{Kroupa2001} initial mass function. Since only a limited number of spaxels observed with the BH2 grating have sufficient S/N to reliably measure the stellar velocity, we use the Voronoi tessellation method ({\tt VorBin}; \citealt{Cappellari2003}) to adaptively rebin the spaxels to a minimum continuum S/N of 20~\AA$^{-1}$ at 4100 \AA. Similarly, we apply {\tt VorBin} to the RH2 channel to adaptively bin spaxels with weak H$\alpha$ emission, achieving a minimum H$\alpha$ flux S/N of $\sim$ 5 in each spatial bin and enabling reliable kinematic measurements. Because the $1\farcs35$-wide slices exceed the seeing, surface brightness gradients across each slice can shift the line centroids along the dispersion direction, introducing apparent velocity offsets of up to $\sim25\ {\rm km\ s^{-1}}$ near bright clumps (the ``slit effect''; \citealt{AgraAmboage2014,Labiano2021}). We correct the BH2 and RH2 velocity fields for this effect following \citet{AgraAmboage2014}, using the light distribution within each slice.

We also use the emission line flux in the RH2 observations to estimate star formation rates, gas-phase metallicity, and nitrogen abundance (Section~\ref{sec:gas_abundance}), through using H$\alpha$, [N {\sc ii}] $\lambda$6584, and [S {\sc ii}] $\lambda \lambda$6717, 6731. The integrated H$\alpha$ flux yields a global ${\rm SFR}_{\rm H\alpha}=0.052\pm0.005\ M_\odot\ {\rm yr^{-1}}$ \citep{Kennicutt2012}, in close agreement with the NUV-based estimate above. Because [N~{\sc ii}] and [S~{\sc ii}] emission are usually weak in individual spaxels of the diffuse component, we use {\tt VorBin} to adaptively bin neighboring spaxels to a minimum H$\alpha$ S/N of $\sim50$ per bin, enabling reliable abundance measurements.

We perform the stellar population analysis using the BL-grating observations in two steps. First, we determine the stellar and gas kinematics, and then fit the stellar population with regularization while fixing the kinematic parameters to the best-fitting values from the first step (e.g., \citealt{Tang2022}). To account for nebular continuum emission in the stellar population fit, we generate a separate set of SSP models using {\tt FSPS} \citep{Conroy2009}, adopting the E-MILES stellar spectral library \citep{Vazdekis2016}, MIST isochrones \citep{Choi2016}, and the \cite{Kroupa2001} initial mass function. These models span age bins from 0.001 Gyr to 12.6 Gyr, and metallicity bins from [M/H] = $-1.98$ to $-0.20$ dex. We use the fitted SSP weights to reconstruct the SFHs and derive stellar population properties and stellar masses for the diffuse stellar body and for each of the four brightest clumps.

For the diffuse stellar component of B122, we derive a mass-weighted stellar age of $0.87 \pm 0.20$ Gyr and a metallicity of $[{\rm M/H}]=-1.26\pm0.18$ dex. The $g$-band mass-to-light ratio is 0.22 $M_\odot/L_{\odot, g}$. Combined with the $g$-band magnitude of 18.48, we estimate a stellar mass of $(1.0\pm0.2) \times 10^8\ M_\odot$ for the diffuse stellar body. We find that all the clumps have formed recently, with similar ages of $\sim$ 100--200 Myr. Their stellar masses are estimated to be $\sim10^6\ M_\odot$ each, based on the mass-to-light ratios derived by {\tt pPXF} and their luminosities. Together, all the clumps contain $\sim7\times 10^6\ M_\odot$ in stars, indicating $\sim$ 7\% of the stellar mass of the galaxy is contained in the star-forming clumps. The properties of the four brightest clumps are listed in Table~\ref{tab:clumps}. We present a more detailed analysis and discussion of their stellar populations and SFHs in Section \ref{sec:stellar_pop}.

Throughout this work, we correct for Galactic extinction but not for internal dust attenuation, as the measured Balmer decrement (H$\beta$/H$\gamma \approx 2.18$) is consistent with dust-free Case B recombination (2.13; \citealt{Osterbrock2006}).

\begin{figure*}
\centering
\includegraphics[width=0.835\textwidth]{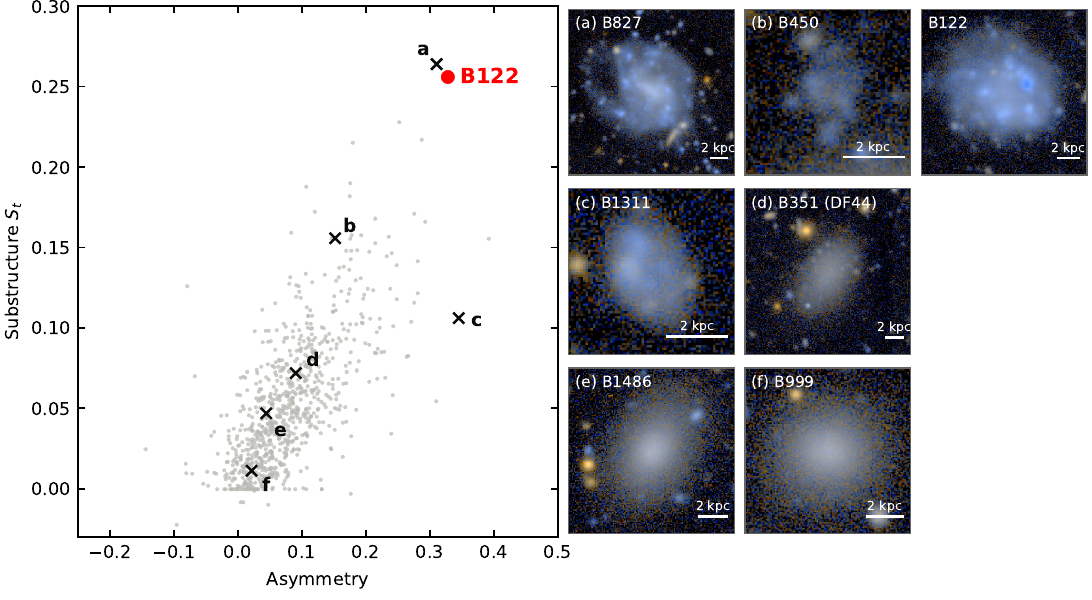}
\caption{The distribution of asymmetry and substructure (St) of Coma LSB dwarfs with $R_{\rm e}>1.25$ kpc in the \cite{Bautista2023} catalog, measured with {\tt statmorph-lsst}. B122 stands out as an extreme outlier, exhibiting both unusually high asymmetry and clumpiness compared with the broader LSB dwarf population. Other representative systems are highlighted for comparison, including the GC-rich UDG DF44. The corresponding galaxy images are shown in the right panels, illustrating the variation from smooth systems with low substructure and/or asymmetry values to increasingly clumpy morphologies.}
\label{fig:statmorph}
\end{figure*}

\section{Results} \label{sec:results}

\begin{figure*}
\centering
\includegraphics[width=0.85\textwidth]{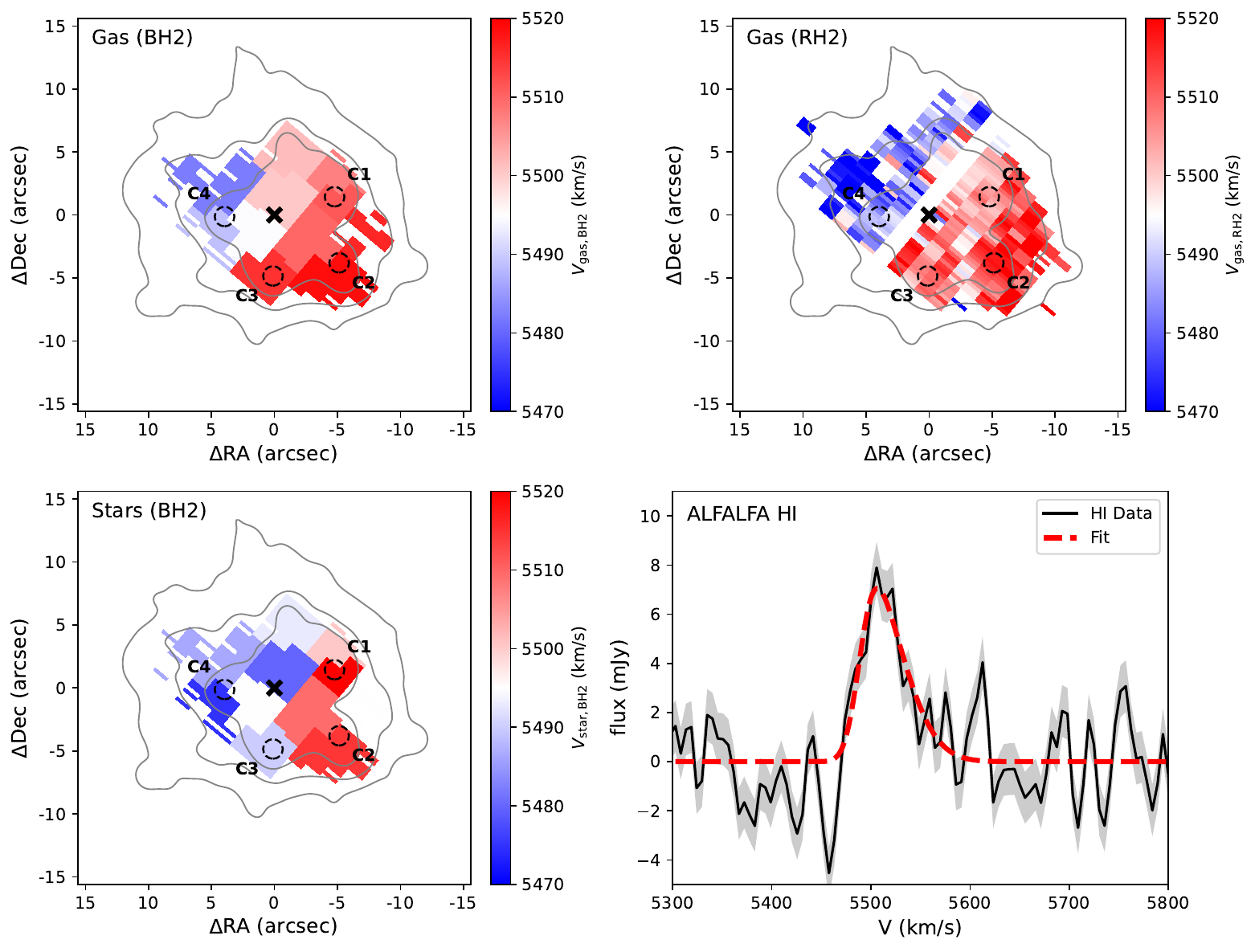}
\caption{Kinematics of B122, with contours of galaxy light in $g$-band overplotted on the velocity maps. {\em Top row}: line-of-sight velocity fields of the ionized gas measured from the BH2 and RH2 grating observations, respectively. {\em Bottom left}: stellar velocity field from the BH2 grating observation. In all three maps, the black cross marks the center of the diffuse stellar body, and the brightest clumps C1--C4 are marked with small circles. The stellar and ionized-gas components both exhibit consistent, ordered rotation. {\em Bottom right}: integrated H~{\sc i} spectrum of B122 from ALFALFA (black), with the $1\sigma$ uncertainty shown by the gray shaded band. The red dashed curve shows the best-fitting skewed Gaussian. The H~{\sc i} systemic velocity is consistent with those measured from the stellar and ionized-gas components.}
\label{fig:kinematics}
\end{figure*}

\begin{figure*}
\centering
\includegraphics[width=0.65\textwidth]{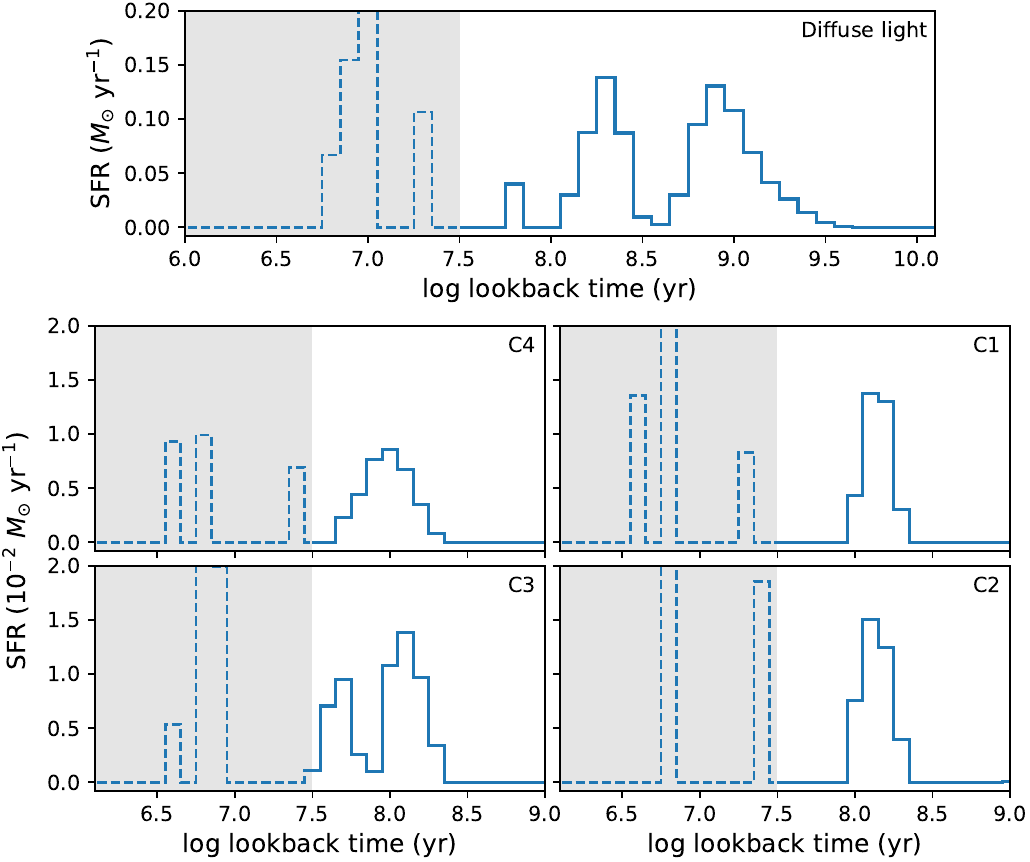}
\caption{SFHs of the diffuse stellar body of B122 ({\em top}) and of the four brightest clumps ({\em middle and bottom, labeled in the panels}). For the diffuse component, about 12.5\% of the mass was formed within the last 300 Myr. All of the clumps were formed during a $\sim$ 100--200 Myr SFR peak in lookback time, with some clump-to-clump temporal variations. Dashed lines on a gray background mark lookback times below $10^{7.5}$ yr, where the SFHs are not well-constrained by our spectra.}
\label{fig:sfh}
\end{figure*}

\subsection{B122 Uniqueness}
\label{sec:B122_uniqueness}

We present various properties of B122 in Table \ref{tab:properties}. B122 has a low SB ($\mu_{g,0}=23.93$ mag arcsec${}^{-2}$) and a large size ($R_{\rm e}=3.2$ kpc). Although it does not strictly satisfy the conventional UDG definition of \cite{vanDokkum2015} based on central SB, its stellar surface density is comparably low, owing to its young stellar population. Passive fading would therefore bring its surface brightness into the formal UDG regime. There are no obvious nearby companions or massive galaxies in the vicinity of B122. Compared with the Coma LSB dwarf population cataloged by \citet{Bautista2023}, B122 is among the largest, most luminous, and bluest systems, with a very shallow light profile (S\'ersic $n=0.52$, compared with a typical $n\approx0.9$). Its global SFR of about $0.05$--$0.06\ M_\odot\ {\rm yr^{-1}}$ is high for its stellar mass, lying 0.6 dex above the star-forming main sequence of nearby dwarfs ($\approx0.015\ M_\odot\ {\rm yr^{-1}}$; \citealt{Kado-Fong2024}), and nearly 1 dex higher than the typical value for star-forming UDGs of similar stellar mass ($\approx0.006\ M_\odot\ {\rm yr^{-1}}$; \citealt{Zaritsky2026,Vanzanella2026}).

The most massive clumps are estimated to have a stellar mass of $\sim 1.2 \times 10^6\ M_\odot$. These masses are comparable to those of the most massive super star clusters known in nearby dwarf galaxies (e.g., \citealt{Anders2004,Adamo2010}), and are unusually high for compact star-forming regions in low-mass galaxies. Such super star clusters may form in merger-driven starbursts or through in-situ disk instabilities and gas accretion \citep{Krumholz2019,Zhang2020,Lassen2025}. More strikingly, B122 hosts $\sim 10$ clumps that together contribute $\sim25\%$ of the galaxy's total luminosity as mentioned in Section \ref{sec:galfit}. This high clump fraction appears unusual among low-mass galaxies in the Local Universe, as well as among nearby star-forming UDGs \citep{Khim2026}, but is more typical for galaxies at $z\sim2$ \citep{Guo2015}.

To further quantify the morphological uniqueness of B122 within the LSB dwarfs, we apply {\tt statmorph-lsst} \citep{Rodriguez-Gomez2019,Sazonova2026} to the HSC imaging of the Coma LSB dwarf sample from \cite{Bautista2023}, selecting galaxies with $R_{\rm e}>1.25$ kpc. We visually remove galaxies that are substantially contaminated by foreground stars or background galaxies. Also it should be kept in mind that the distances for most of this sample are not confirmed and could include foreground and background LSB galaxies (cf.\ Figure~2 of \citealt{Kadowaki2021}).

For this galaxy sample, Figure~\ref{fig:statmorph} shows the distribution of the asymmetry and substructure indices. B122 stands out as an extreme outlier, exhibiting both unusually high asymmetry and substructure compared with the broader LSB dwarf population. This asymmetry is driven by the spatial distribution and varying luminosities of the clumps, together with the lopsided diffuse light. The light profile is sharply truncated toward the Coma cluster center on B122's southwest side (S\'ersic $n\approx0.3$), but more extended to the northeast ($n\approx0.7$), with excess luminosity on the northeast side corresponding to $\sim 6\%$ of the total galaxy light. This pattern appears to represent a bow-shock and tail structure that is shaped by ram pressure. In contrast, the well-studied quiescent UDG DF44 in Coma \citep{vanDokkum2015}, despite hosting a rich GC system, lies near the center of the distribution in Figure~\ref{fig:statmorph} and does not show similarly extreme morphological parameters, as the GCs are much fainter than B122's star-forming clumps. Figure~\ref{fig:statmorph} also shows a few representative galaxies at different locations in the distribution, where higher substructure and/or asymmetry values tend to correspond to more clumpy morphologies. Several systems such as B827 show visible clumpy structures, but none hosts clumps as luminous as those in B122.

\subsection{Kinematics}
\label{sec:kinematics}

B122 was detected in the Arecibo Legacy Fast ALFA Survey (ALFALFA; \citealt{Haynes2018}), showing an H~{\sc i} mass of $10^{8.95}\ M_\odot$ (assuming a distance of 100 Mpc), which is typical for both classical dwarfs and UDGs at this stellar mass \citep{Maddox2015,Leisman2017,Zhou2026}. Figure \ref{fig:kinematics} shows the H~{\sc i} velocity profile of B122 from ALFALFA, with reported widths of $W_{50} = 45 \pm 9\ {\rm km\ s^{-1}}$ and $W_{20} = 75\ {\rm km\ s^{-1}}$. Using $W_{50}/2$ as a proxy for the projected rotation speed, and an inclination of $\sim37^\circ$ derived from B122's axis ratio ($b/a=0.77$) while assuming the intrinsic shapes of luminous dwarf irregulars \citep{Roychowdhury2013}, we estimate a deprojected $V_{\max}$ of $\sim 40\ {\rm km\ s^{-1}}$. Combining this value with an H~{\sc i} radius $R_{\rm HI}\sim 8.6$ kpc inferred from the empirical H {\sc i} size--mass relation \citep{Wang2016} gives a dynamical mass of $\sim3\times10^9\ M_\odot$ within $R_{\rm HI}$, about 2.5 times the baryonic mass. Although this ratio implies the presence of dark matter, it is lower than that of typical dwarfs of similar baryonic mass \citep{Lelli2019} and is comparable to the ratios found in H~{\sc i}-rich UDGs that lie below the baryonic Tully--Fisher relation and exhibit high baryon fractions \citep{ManceraPina2019,ManceraPina2020}. However, stronger conclusions will require spatially resolved H~{\sc i} observations to more directly measure $R_{\rm HI}$ and the inclination.

The H~{\sc i} velocity profile exhibits a clear asymmetry, with an extended tail toward the redshifted side. Following the definition of \cite{Haynes1998}, we quantify this asymmetry as $1.48 \pm 0.18$, defined as the ratio of the integrated fluxes in the two halves of the profile, which is significantly greater than 1. This asymmetric profile suggests that the H~{\sc i} gas is moderately disturbed, potentially as a result of environmental interactions such as RPS in Coma. The overall H~{\sc i} velocity ($5510 \pm 5\ {\rm km\ s^{-1}}$) is consistent with that of the stars ($5495 \pm 4\ {\rm km\ s^{-1}}$) and with the ionized gas ($5507 \pm 2\ {\rm km\ s^{-1}}$).

Figure \ref{fig:kinematics} shows the stellar and ionized-gas velocity fields of B122, both of which exhibit a clear signature of ordered rotation. The four brightest clumps have velocities within $7\ {\rm km\ s^{-1}}$ of the surrounding ionized-gas velocity field interpolated to their positions, indicating that they follow the same rotation. 
Fitting a sinusoid to the ionized-gas and star velocities of the Voronoi bins in Figure~\ref{fig:kinematics} as a function of position angle gives a projected rotation amplitude of $\sim22\ {\rm km\ s^{-1}}$, along with a $\sim 30^\circ$ twist (for the gas) relative to the isophotes, suggesting mild deviations from equilibrium. This is consistent with the $W_{50}/2$ of H~{\sc i} component, yielding the same $V_{\max}\sim40\ {\rm km\ s^{-1}}$ after correcting for inclination. The global ionized-gas velocity dispersion $\sigma$ is $20~{\rm km\ s^{-1}}$, yielding $V_{\rm max}/\sigma\sim2$. This ratio indicates that ordered rotation provides a dynamically significant contribution to the support of B122. Such kinematics are consistent with the presence of a dynamically important disk component and are broadly typical of gas-rich, star-forming dwarf galaxies \citep{Moiseev2015}. The presence of an extended rotationally supported component is also consistent with observations of gas-rich UDGs in low-density environments \citep{ManceraPina2020}.

\subsection{Stellar populations}
\label{sec:stellar_pop}

Figure \ref{fig:sfh} shows the SFHs of the diffuse stellar body and the four brightest clumps of B122, recovered from our {\tt pPXF} spectral fits. The diffuse component is young and metal-poor, with a mass-weighted stellar population age of $0.87 \pm 0.20$ Gyr and a metallicity of [M/H] = $-1.26 \pm 0.18$ dex. Compared to the \cite{Kirby2013} stellar mass--metallicity relation and to other UDGs \citep{FerreMateu2018,Rong2020}, B122 is slightly metal-poor but remains within the scatter. The SFH does not show a single burst or smoothly declining history. Instead, about 12.5\% of the stellar mass formed during an extended episode of enhanced star formation over the last $\sim$ 300 Myr, on top of an old underlying population that dominates the total mass budget. The recent SFH shows a prominent peak in star formation $\sim 200$ Myr ago. The detailed SFH of the older population is not well-constrained by our data.

For the clumps, we find concentrated episodes of recent star formation that are broadly synchronized with the diffuse component, but perhaps slightly delayed, with the clump activity peaking $\sim$ 100--200 Myr ago. We do not find evidence for a substantial old stellar population within the clumps, suggesting that they are not pre-existing structures but formed completely during the recent starburst.

Although the overall SFHs are similar, we find differences among the clumps that appear to correlate with their spatial locations. As shown in Figure \ref{fig:sfh}, the main SFH peaks of the two clumps in the east occurred $\sim$ 50--100 Myr more recently than those of the two clumps in the west. The SFHs at lookback times below $10^{7.5}$ yr are not well-constrained by our spectra, as the optical continua of very young populations are nearly featureless and Balmer absorption is filled in by nebular emission, preventing reliable separation of their contributions across age bins. We therefore exclude this regime from our SFH interpretation and instead use emission lines to trace the most recent star formation. Interestingly, the two clumps in the west currently exhibit substantially stronger star formation than the other clumps, as indicated by their emission line strengths (Figures \ref{fig:luminosity_map} and \ref{fig:spectra}). The brightest clump C1 even appears to exhibit a Wolf--Rayet feature with a blue bump around He {\sc ii} $\lambda$4686 \citep{Conti1991}, indicating the presence of young, massive stars and very recent star formation ($\sim$ 3--6 Myr; \citealt{Leitherer1999}). Given that the clumps have already assembled substantial stellar masses over the past $\sim300$~Myr, their continued star formation suggests an extended period of activity, rather than a single burst followed by rapid suppression from stellar feedback. This suggests that the clumps have sustained gas supply, or that there is an external mechanism weakening the disruptive impact of stellar feedback.

\subsection{Gas abundance}
\label{sec:gas_abundance}

\begin{figure}
\centering
\includegraphics[width=0.472\textwidth]{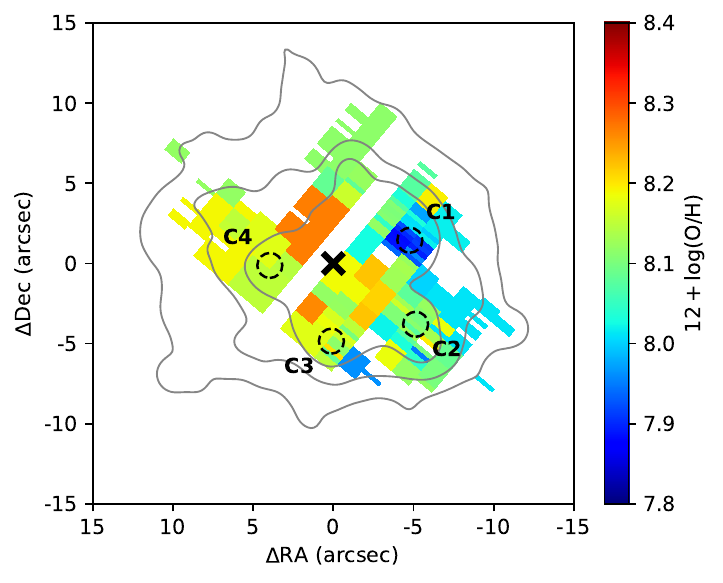}
\caption{Map of gas-phase metallicity derived using N2 indicator, with contours of galaxy light in $g$-band overplotted, and brightest clumps marked with small circles. Only the spatial bins with ${\rm S/N}>3$ in [N {\sc ii}] $\lambda$6584 emission line fluxes are plotted. The black cross marks the center of B122. The western half of B122 is slightly more metal-poor than the eastern half by $\sim0.1$ dex, while the brightest clump C1 shows a significantly lower metallicity of $\sim0.2$ dex below the global value, which may suggest that the most recent star formation is fueled by metal-poor gas inflow.}
\label{fig:gas_metallicity}
\end{figure}

Figure \ref{fig:gas_metallicity} shows the spatially resolved gas-phase metallicity map, derived from the N2 indicator ([N {\sc ii}] $\lambda$6584/H$\alpha$) using the calibration by \cite{Marino2013}. B122 has an overall gas-phase metallicity of $12+\log({\rm O/H})=8.1 \pm 0.1$, consistent with the general mass--metallicity relation of dwarfs \citep{Scholte2024,McQuinn2026}, and by extension, also consistent with the average star-forming UDG population which are thought to follow the standard dwarf relation \citep{Rong2020}. The gas has an average negative radial gradient of metallicity. The western half of B122 appears slightly more metal-poor than the eastern half by $\sim 0.1$ dex, while the brightest clump C1 shows a significantly lower metallicity of $\sim 0.2$ dex below the global value. Among the spaxels overall, there is a strong anti-correlation between metallicity and SFR (cf.\ \citealt{Long2025,Lassen2025}). These differences may suggest that the most recent star formation is fueled by the inflow of metal-poor gas (e.g., \citealt{Tang2022,Cosens2024,Long2025}). The emission line flux ratios do not show any evidence of AGN activity or shocks.

\begin{figure}
\centering
\includegraphics[width=0.472\textwidth]{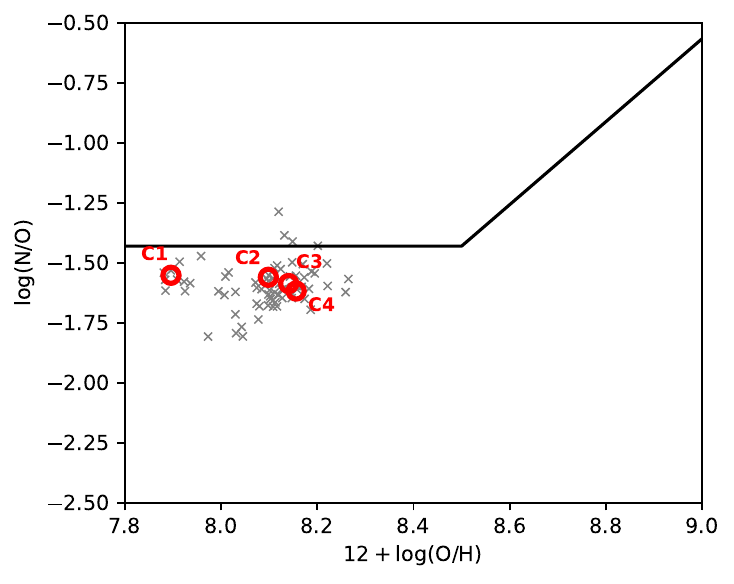}
\caption{The distribution of nitrogen abundance N/O and gas-phase metallicity for the spatial bins created in Section \ref{sec:gas_abundance}, shown as gray crosses. Only the spatial bins with ${\rm S/N}>3$ in [N {\sc ii}] $\lambda$6584, and in [S {\sc ii}] $\lambda \lambda$6717, 6731 emission line fluxes are plotted. Red circles mark the four brightest clumps. The black line shows the best-fit N/O enrichment model from \cite{Andrews2013}.}
\label{fig:nitrogen_abundance}
\end{figure}

Figure \ref{fig:nitrogen_abundance} places the individual spatial bins and the four brightest clumps on the nitrogen abundance log(N/O) versus gas-phase metallicity diagram. Here, we estimate N/O using the N2S2 index ([N {\sc ii}] $\lambda$6584/[S {\sc ii}] $\lambda \lambda$6717, 6731) following the calibration of \cite{Perez-Montero2021}. B122 has a relatively low $\log({\rm N/O}) \sim -1.55$, lying $\sim$ 0.1 dex below the primary nitrogen abundance plateau in \cite{Andrews2013}. This offset should be interpreted cautiously, as our N2S2-based estimates are being compared with a direct-method relation, and in any case, B122 is within the scatter of the N/O distribution for low-metallicity galaxies \citep{Perez-Montero2021}. The four brightest clumps lie within the region traced by the diffuse gas, with no significant difference in $\log({\rm N/O})$ at fixed 12+log(O/H). Unlike the gas-phase metallicity, N/O shows no clear spatial variation across the galaxy. The combination of a spatial O/H variation and a nearly uniform N/O is consistent with dilution by metal-poor gas inflow, which can lower O/H while leaving N/O largely unchanged \citep{Koppen2005}.

\section{Discussion} \label{sec:discussion}

\subsection{The trigger of the starburst episode $\sim$200 Myr ago}

In Section \ref{sec:intro}, we suggested dwarf--dwarf interactions and ram pressure as possible triggers of B122's starburst. One possibility is a high-speed ``bullet-dwarf'' collision, in which shocked gas separates from collisionless dark matter and forms a diffuse, dark-matter-deficient remnant hosting massive star clusters \citep{Silk2019,Lee2021,vanDokkum2022}. However, 
the dynamical mass estimate of B122 implies the presence of dark matter (Section~\ref{sec:kinematics}). Furthermore, the dominant older stellar population ($\sim87.5$\% of the stellar mass formed before the recent episode) also disfavors this scenario, in which most stars should form from the collision-shocked gas \citep{vanDokkum2022}. 

Other strongly disruptive mergers are also disfavored by the dynamical timescale. With $R_{\rm e}=3.2$ kpc and $V\sim40\ {\rm km\ s^{-1}}$, B122 has a crossing time of $t_{\rm cross}\sim80$ Myr. A merger associated with the starburst $\sim200$ Myr ago would have occurred only $\sim2$ crossing times ago, and should have left more obvious disturbances (e.g., \citealt{Zhang2020b}). B122's coherent stellar and ionized-gas rotation instead suggests that its disk was largely preserved.

Milder interactions, such as tidal encounters with nearby massive galaxies, may occur without significantly disturbing the galaxy's kinematics \citep{Rossi2026}. However, B122 has no obvious massive companions in its vicinity. A close encounter with a dark or nearly-dark matter subhalo offers another gravitational trigger, potentially compressing gas and enhancing star formation without a visible companion while perturbing gas and stars \citep{Starkenburg2016a,Starkenburg2016b}. In some cases, simulations show that such encounters might preserve ordered gas rotation and sustain enhanced star formation over extended periods \citep{Starkenburg2016a}. B122's coherent rotation and extended SFH therefore do not rule out this scenario.

However, several pieces of evidence point toward an RPS-related scenario. B122's diffuse light is truncated toward the Coma cluster center and extended in the opposite direction, while its H~{\sc i} velocity profile is asymmetric, suggesting ongoing environmental interaction. These may indicate that the galaxy is compressed on the leading side and displaced downstream by ram pressure \citep{Abramson2011,Matijevic2026}\footnote{RPS acts in the direction of the galaxy orbit, which is not likely to be purely radial, particularly when seen in projection. Nevertheless, RPS tails preferentially point away from the cluster center \citep{Salinas2024}.}. The phase-space position supports recent infall. B122 lies in the outskirts of the Coma cluster, at a projected clustercentric radius of $0.8\ R_{\rm vir}$, where $R_{\rm vir}=2.9$ Mpc \citep{Lokas2003}. Its line-of-sight velocity is offset from the Coma systemic velocity by $\sim1350\ {\rm km\ s^{-1}}$, or $\sim1.25$ times the cluster velocity dispersion \citep{Colless1996}. The location and velocity are consistent with recent infall \citep{Rhee2017} and may suggest relatively weak RPS \citep{Jaffe2018,Wang2020}. The preserved disk kinematics additionally suggests that B122 has not experienced substantial dynamical disturbance, consistent with an early stage of infall. Furthermore, B122 is one of only three cases from the \citet{Bautista2023} sample of Coma LSB dwarfs to have an ALFALFA detection, suggesting that it is one of the most recent infallers, without yet experiencing substantial gas loss from the environment.

Observations show that Coma stripping candidates exhibit elevated current star formation \citep{Roberts2020}, while low-mass post-starburst galaxies show evidence of substantial activity before quenching \citep{Grishin2026}. Post-starburst dwarfs affected by RPS in Coma formed 10--60\% of their stellar mass in a final burst \citep{Grishin2021}. B122 formed $\sim12.5\%$ of its stellar mass over the past $\sim300$ Myr and may represent an earlier stage of the same evolutionary sequence. Simulations also suggest that ram pressure can enhance star formation through gas compression \citep{Steyrleithner2020} or redistribution that increases the central dense-gas fraction \citep{Zhu2024}. This provides a plausible explanation for B122's enhanced star formation in the last $\sim$~300 Myr along with a signature of metal-poor gas inflow. The clumps may also sustain their star formation through continued gas inflow driven by ram pressure. \cite{Grishin2026} suggested that galaxies infalling on high-angular-momentum orbits may sustain substantial star formation before gas stripping, as ram pressure increases more gradually along such orbits. This scenario may apply to B122.

Although rapid ram-pressure gas removal can expand a dwarf galaxy's stellar disk and subsequently lower their surface brightness through quenching and fading \citep{Grishin2021}, B122 remains gas rich and shows no clear signatures of strong stripping. Its diffuse morphology may therefore predate the current interaction, suggesting that B122 was already a UDG before infalling into Coma, although its pre-infall size remains unknown. The very low S\'ersic index of the stars does suggest there may have been movement of central material outward, perhaps starting with a compact ring of clumps as in NGC~1522 \citep{Long2025} that migrated out to 2~kpc. B122 may represent a transitional stage in the environmental evolution of a gas-rich diffuse dwarf, with continued ram-pressure stripping eventually depleting its gas and quenching star formation. Such a stage may be short-lived, consistent with the rarity of star-forming UDGs in clusters \citep{Singh2019,Junais2022,Rossi2026}.

The formation of B122's unusually luminous clumps nevertheless requires sufficiently strong local gas collapse. Thick disks in nearby dwarfs can suppress large-scale gravitational instabilities while still allowing local gas collapse and star formation \citep{Elmegreen2015}. H~{\sc i}-rich UDGs can exhibit low star-formation efficiencies despite substantial gas reservoirs \citep{Kado-Fong2022b}, indicating that gas content alone cannot explain B122's unusually luminous clumps. Ram pressure may enhance local collapse through gas compression, although internal processes and past encounters could also contribute. Resolved cold-gas observations are needed to assess disk stability and distinguish between these mechanisms.

\subsection{The survival and recurrent activity of the clumps}

The clumps show substantial star-forming activity starting $\sim300$~Myr ago, while strong emission lines and possible Wolf--Rayet features indicate much younger stellar populations (Figures~\ref{fig:spectra} and \ref{fig:sfh}). Their extended star formation period raises questions about their survival, as unbound stellar groups should disperse over their internal crossing times, while rapid gas removal can disrupt initially bound systems \citep{PortegiesZwart2010}. In any case, the clumps can be interpreted as star-forming complexes containing bound clusters embedded in diffuse stellar populations, with the densest structures surviving gas removal \citep{Krumholz2019}.

In the RPS scenario, gas compression could trigger successive episodes of star formation. Feedback may disperse dense gas and temporarily suppress star formation \citep{Chevance2020}, while continued compression and gas inflow could fuel subsequent episodes. Such recurrent activity, together with surviving bound clusters, may help maintain the clumps' prominence despite the dispersal of unbound stars. B122's substantial gas reservoir could sustain this activity. 

The west clumps exhibit earlier SFH peaks but stronger current star formation than their eastern counterparts, resembling the enhanced leading-side activity observed in RPS galaxies \citep{Steyrleithner2020,Roberts2022}, on the same side where the diffuse light is truncated. This asymmetry may reflect uneven gas compression or supply, with gas-phase metallicity variations suggesting possible metal-poor inflow into the western clumps. However, the clumps rotate within the disk, so their present positions may not coincide with the sites of earlier starbursts. The $\sim80$~Myr dynamical timescale is comparable to the inferred SFH differences, suggesting that the clumps may have experienced compression at different orbital phases.

The clumps form an unusual ring-like configuration at $\sim2$~kpc, with no prominent clumps near the galaxy center, unlike typical ``jellyfish'' galaxies, where star-forming clumps are concentrated in one-sided tails and/or central regions \citep{Poggianti2019,Grishin2021}. The northeastern extension also lacks an obvious H$\alpha$ counterpart. Similar ring configurations are found in NGC~1522 \citep{Long2025} and the collisional system Arp~147 \citep{Randriamanakoto2026}. The ring's origin remains unclear, with possible explanations including an encounter with a purely dark matter subhalo as discussed above, or ram-pressure-driven angular momentum transport \citep{Schulz2001}. Current observations do not distinguish between these scenarios.

\subsection{Are the Clumps GC progenitors?}

B122's brightest clumps have estimated stellar masses of $\sim 10^6\ M_\odot$, comparable to massive GCs in dwarf galaxies. They contribute a total of $\sim7\%$ of the stellar mass and $\sim25\%$ of the $g$-band light, which we have found is unusual for UDGs in Coma (Section~\ref{sec:B122_uniqueness}) and is likely unusual for the population of gas-rich UDGs in general. For example, although the large KCWI survey of UDGs by \cite{Khim2026} turned up many examples of emission-line clusters, these are much fainter than the clumps in B122. This raises the possibility that the clumps in B122 may mark the transformation of an ordinary dwarf into a GC-rich UDG, in an event that is rare today, but was more common in the early Universe \citep{Peng2016,FerreMateu2023}.

Determining whether or not B122's clumps can survive and eventually evolve into GCs requires resolving their internal structure, which is beyond the spatial resolution of our current data. Indirect clues can instead be obtained from nearby low-mass galaxies hosting clumps that have been resolved at higher spatial resolution. \cite{Buzzo2025c} identified YMCs with masses of $\sim10^5\ M_\odot$ within clumps in the UDG GAMA~526784, which they interpreted as products of dwarf--dwarf interactions. Similarly, \cite{Fahrion2026} found that the clumps in the jellyfish galaxy NGC~1427A, currently infalling into the Fornax cluster, host multiple YMCs with masses of $\sim10^{5.5}\ M_\odot$. These observations show that massive star-forming clumps are not monolithic structures, but can instead host different substructures, which has also been observed in more massive galaxies (e.g., \citealt{Whitmore1999,Randriamanakoto2026}).

An order-of-magnitude estimate suggests that the B122 clumps could produce massive young star clusters. For a clump of $\sim 10^6\ M_\odot$, we can assume a high cluster formation efficiency of $\Gamma \sim50\%$ (i.e., fraction of new stars formed in bound clusters), representative of strongly star-forming environments in ram pressure \citep{Kruijssen2012,Adamo2020,Randriamanakoto2026}. Adopting a $M^{-2}$ power-law initial cluster mass function \citep{PortegiesZwart2010} and normalizing the high-mass tail to one cluster \citep{Elmegreen2018}, we obtain $M_{\max}[1+\ln(M_{\max}/M_{\min})] \sim \Gamma M_\star$. For $M_{\min}=100\ M_\odot$, this yields $M_{\max}\sim10^{4.8}\ M_\odot$. An independent empirical relation between the Schechter truncation mass $M_{\rm c}$ and the global SFR surface density \citep{Johnson2017} gives $M_{\rm max}\sim10^{4.2}\ M_\odot$ for B122, given its $\Sigma_{\rm SFR} \sim 10^{-2.5}\ M_\odot\ {\rm yr^{-1}\ kpc^{-2}}$. However, this relation is calibrated using galaxy-wide averages, whereas B122's star formation is concentrated in its clumps. The higher local $\Sigma_{\rm SFR}$ ($\gtrsim 10^{-1.5}\ M_\odot\ {\rm yr^{-1}\ kpc^{-2}}$, see Figure~\ref{fig:luminosity_map}) may therefore permit a larger $M_c$.

The gas-phase nitrogen abundance provides an additional test of the clumps as GC progenitors. Young massive and super star clusters have been observed to produce localized nitrogen enhancements in their surrounding ionized gas (e.g., \citealt{Smith2016,Pascale2023}). We do not detect a comparable excess in B122's clumps. However, this non-detection does not rule out localized enrichment, as the enhanced nitrogen in these systems is confined to scales of tens of parsecs and could be strongly diluted at our spatial resolution ($\sim$ 500 pc; cf.\ \citealt{Long2025}). Notably, the brightest clump shows the Wolf--Rayet feature, confirming the presence of very massive stars capable of producing such enrichment. Spectroscopic observations with higher spatial resolution would provide a more sensitive test of localized chemical self-enrichment.

For B122's clumps to evolve into GCs, their constituent clusters must remain bound through gas removal, stellar mass loss, and subsequent dynamical evolution \citep{Krumholz2019}. Simulations of dwarf galaxies suggest that surviving clusters could retain $\sim40\%$ of their birth mass after 8 Gyr, including stellar evolution and dynamical losses \citep{Shao2021}, implying a remaining mass of $\sim10^{4.6}\ M_\odot$ for a cluster born with $10^5\ M_\odot$, comparable to many GCs. While massive clumps can undergo substantial tidal stripping \citep{Bournaud2014}, simulations show that sufficiently dense clusters can survive in low-mass galaxy potentials \citep{Penarrubia2009}, and of course GCs are observed to exist pervasively in dwarf galaxies. B122's low stellar surface density may favor YMC survival \citep{Moreno-Hilario2024}, making its clumps promising sites for forming long-lived GCs. However, their unresolved internal structures prevent us from determining whether or not they contain sufficiently dense, bound clusters. Higher-resolution observations resolving individual clusters and measuring their sizes, masses, and ages are needed to assess their long-term survival.

\section{Conclusion} \label{sec:conclusion}

In this work, we have conducted a comprehensive analysis of an unusual star-forming UDG in the Coma cluster, B122, and its bright clumps using Subaru/HSC imaging and new Keck/KCWI spectroscopy. Our main findings and conclusions are as follows:

\begin{itemize}

\item
B122 hosts $\sim$ 10 luminous star-forming clumps that together contribute $\sim$ 25\% of its $g$-band total luminosity and $\sim$ 7\% of its total stellar mass, making it an extreme outlier among UDGs and other low-redshift dwarf galaxies. The clumps have typical stellar masses of $\sim10^6\ M_\odot$, with the most massive one reaching $\sim1.2\times10^6\ M_\odot$, comparable to the most massive super star clusters known in nearby dwarfs. The clumps form a distinctive ring-like configuration at $\sim2$ kpc from the center of B122.

\item
The diffuse stellar body of B122 is young and metal-poor, with a mass-weighted age of 0.9 Gyr and a metallicity of ${\rm [M/H]}=-1.3$. About 12.5\% of its stellar mass formed in an episode of enhanced star formation over the last $\sim$ 300 Myr. All of the clumps formed within this episode.

\item
B122 shows evidence of weak RPS, including a lopsided diffuse light distribution, truncated on the side facing the Coma center and extended on the opposite side, along with a moderate but significant asymmetry in its H~{\sc i} gas velocity profile. Its recent enhanced star formation, ordered stellar and ionized-gas disks, phase-space position, and cold gas reservoir are consistent with an early stage of infall into the Coma cluster. High-resolution H~{\sc i} observations are essential to decisively test this scenario.

\item
The spatial variations in the SFHs and gas-phase abundances suggest that RPS-driven inflow of metal-poor gas may have triggered the recent starburst and helped sustain the intense star formation in the clumps. However, the origin of the ring-like clump configuration remains unclear.

\item 
It remains unclear whether or not the clusters within these clumps will ultimately evolve into GCs. The individual clump masses should permit the formation of YMCs with  masses $\gtrsim 10^5\ M_\odot$ at birth. We do not detect a clear nitrogen abundance enhancement for ionized gas associated with the clumps, which may result from dilution effects. Higher-resolution imaging and spectroscopy will be critically needed to resolve the internal structure, detailed abundances, and feedback-driven outflows of the clumps, while detailed dynamical modeling will be essential for determining the survivability of their constituent clusters in the tidal field of the host galaxy.

\end{itemize}

The origins of GC-rich UDGs are a long-standing mystery, and B122 appears to provide a ``missing link'' example observed in formation on the outskirts of a high-mass cluster (where GC-rich UDGs are more prevalent).  More thorough investigation of the galaxy's past, present, and future would elucidate its relevance. Models of the ram pressure and clump formation are needed, as well as of the galaxy's long-term evolution in the cluster environment. GC-rich UDGs in clusters have a suite of characteristic properties (round, slow-rotating, positive stellar metallicity gradients) that are seen as indicators of a ``failed galaxy'' formation history. The present-day properties of B122 suggest it may be a different breed of GC-rich UDG, triggered by cluster infall and ram pressure, while other cases may form through internal processes (e.g., \citealt{Danieli2022,Janssens2022,Fielder2023,Buzzo2025c}).

Coma also hosts a larger population of star-forming LSB dwarfs at different stages of RPS, including systems with a range of clumpiness. These galaxies provide a chance to better understand how RPS affects the evolution of clumps, galaxy morphology, and potentially GC formation. Other UDGs at different stages in the evolutionary process could be identified. For example, HST studies of cluster UDGs have focused only on the quiescent cases, while a young, recent infaller in the Perseus cluster was serendipitously included in Euclid imaging and found to host a substantial population of GCs \citep{Tang2026a}. Clumpy LSB systems in other local-Universe environments could also be uncovered using medium- and narrow-band surveys designed to identify emission-line features (e.g., \citealt{Mintz2024}), thus providing a more complete picture of the formation and evolution of these systems.
Finally, if clumpy UDGs like B122 were common in the early Universe, they may be revealed in James Webb Space Telescope observations of these epochs (cf.\ \citealt{Fujimoto2025}). 

\begin{acknowledgments}

This work is supported by National Science Foundation grant AST-2308390. The Observatory was made possible by the generous financial support of the W. M. Keck Foundation. The authors wish to recognize and acknowledge the very significant cultural role and reverence that the summit of Maunakea has always had within the indigenous Hawaiian community. We are most fortunate to have the opportunity to conduct observations from this mountain.

\end{acknowledgments}

\bibliography{reference}{}
\bibliographystyle{aasjournal}

\end{document}